\documentclass[journal]{IEEEtran}
\usepackage[table]{xcolor}
\usepackage{bm}
\usepackage{mathrsfs}
\usepackage{amsmath,amssymb}
\usepackage{graphicx}
\makeatletter
\newcommand*\bigcdot{\mathpalette\bigcdot@{.8}}
\newcommand*\bigcdot@[2]{\mathbin{\vcenter{\hbox{\scalebox{#2}{$\m@th#1\bullet$}}}}}
\makeatother
\usepackage{pgfplots}
\usepackage{algorithm}
\usepackage{algpseudocode} 
\usepackage[colorlinks, linkcolor=red, anchorcolor=blue, citecolor=green]{hyperref}
\usepackage{epstopdf}
\usepackage{color}
\usepackage{cite}
\usepackage[T1]{fontenc}
\usepackage{multirow}
\usepackage{hhline}

\begin{document}

\title{Intelligent Reflection Enabling Technologies for Integrated and Green Internet-of-Everything Beyond 5G: Communication, Sensing, and Security}

\author{Wei~Shi,~\IEEEmembership{Student Member,~IEEE},
       Wei~Xu,~\IEEEmembership{Senior Member,~IEEE},
       Xiaohu~You,~\IEEEmembership{Fellow,~IEEE},
       Chunming~Zhao,~\IEEEmembership{Member,~IEEE},
       and Kejun~Wei\vspace{-15.pt}


\thanks{Wei~Shi, Wei~Xu (corresponding author), Xiaohu~You, and Chunming~Zhao are with the National Mobile Communications Research Laboratory, and Frontiers Science Center for Mobile Information Communication and Security Southeast University, Nanjing 210096, China, and are also with the Purple Mountain Laboratories, Nanjing 211111, China (e-mail:\{wshi, wxu, xhyou, cmzhao\}@seu.edu.cn).}

\thanks{Kejun~Wei is with the China Academy of Information and Communications
Technology (e-mail: weikejun@caict.ac.cn).} 
}

\maketitle

\begin{abstract}
Internet-of-Everything (IoE) has gradually been recognized as an integral part of future wireless networks. In IoE, there can be an ultra-massive number of smart devices of various types to be served, imposing multi-dimensional requirements on wireless communication, sensing, and security. In this article, we provide a tutorial overview of the promising intelligent reflection communication (IRC) technologies, including reconfigurable intelligent surface (RIS) and ambient backscatter communication (AmBC), to support the requirements of IoE applications beyond the fifth-generation (5G) wireless communication network. Specifically, we elaborate on the benefits of IRC-assisted IoE in the context of the space-air-ground integrated communications and green communications, which are regarded as key features of supporting future IoE application from society and industries. Furthermore, we envision that the IRC-assisted communication and sensing can mutually benefit each other and articulate multiple ways of enhancing the security in IoE by the IRC. Numerical results help illustrate the importance of the IRC in unfavorable secrecy environments. Finally, open research issues and challenges about the IRC-assisted IoE are presented.
\end{abstract}

\IEEEpeerreviewmaketitle

\vspace{-12.pt}
\section{Introduction}

With the world-wide commercial deployment of the fifth-generation (5G) mobile communication systems starting in the mid-2019, research efforts on future wireless networks, e.g., the sixth-generation (6G) networks, are gaining momentum. Especially in 2020, China, the United States, Europe, Japan and South Korea all announced their national research plans for future wireless networks. Even though it still takes time to reach a broad consensus on key technologies, Internet-of-Everything (IoE) has been recognized as one major vision of the future network in both academia and industry \cite{1}.

As forecasted by multiple bodies, applications like smart industry, smart transportation, and smart health are most likely to be fast-growing areas in future network. With these emerging vertical applications, the number of intelligent devices worldwide is expected to experience an explosive growth, which is predicted to reach up to 125 billion by 2030 \cite{2}. Coming with the ultra-massive connectivity of intelligent devices, stringent requirements, e.g., transmission capacity, coverage, latency, reliability, and energy consumption, emerge to realize ultra-massive machine-type communication (umMTC) in IoE. Besides the realization of umMTC, IoE also faces challenges in terms of sensing. Intelligent devices can be distributed over a large area covering indoors, urbans, factories and remote rural regions. It is challenging to obtain and share accurate sensing information of the surrounding environment in these areas, especially under power constraints. Furthermore, sensing information obtained by these intelligent devices is expected to assist the wireless communications via beamforming and channel estimation, referred to as integrated sensing and communication (ISAC), which makes the issue more pressing. A third feature of the IoE is that information security needs careful consideration because the devices are applied in various fields, such as medical care, social life, and even national power grids. In these fields, the acceptance and widespread of future IoE mainly relies on the protection of privacy and security. However, these requirements cannot be completely fulfilled by using existing technologies in the 5G network. It calls for new wireless technologies to support these requirements in future IoE systems beyond 5G.

\begin{figure*}[!t]
  \centering
  \includegraphics[width = 18cm,height= 18cm]{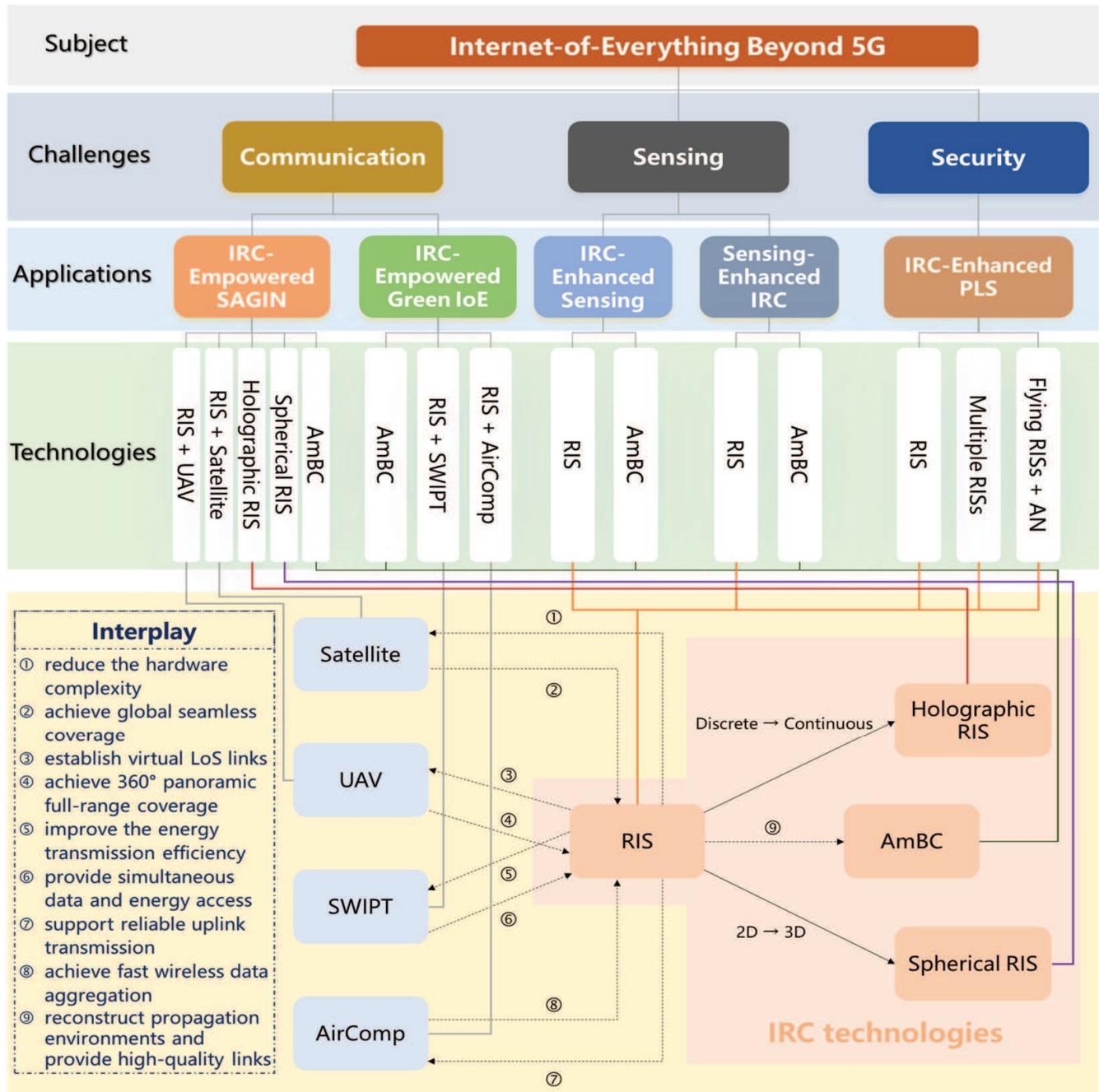}
  \caption{The organization of this paper and the interplay of IRC technologies and other technologies.}
  \label{fig1}
\end{figure*}

Recently, wireless technologies enabled by intelligent reflection communication (IRC), consisting of reconfigurable intelligent surface (RIS) and ambient backscatter communication (AmBC), have received much attention due to their ability to realize smart propagation with low costs and reduced power consumption. Unlike traditional technologies, IRC can actively provide supplementary wireless links, and even transmit additional information, by exploiting environmental signals, which exhibits great potential in IoE applications.

Specifically, RIS is a planar surface composed of a large number of passive reconfigurable reflecting elements \cite{3}, each being able to control the amplitude and phase shift of incident signals independently, thereby collaboratively realizing accurate three-dimensional (3D) beamforming. In this way, the RIS is able to adjust the transmission environment around the IoE devices and strengthens the received signals at the intended receiver for ubiquitous communication. It successfully extends the coverage of devices at cell edges or blocked by obstacles.

As another important component of IRC, AmBC allows intelligent devices to communicate with each other utilizing signals from environmental radio-frequency (RF) sources \cite{4}. Different from changing propagation environments by RIS, AmBC sends data to a backscatter receiver by modulating and reflecting surrounding environmental signals. Thus, the AmBC can be used to connect IoE devices without batteries, build smart sensors, and allow devices to operate independently with minimal manual intervention. Therefore, AmBC serves as a key enabler for achieving high energy efficiency in future IoE with marginal maintenance costs. 

Furthermore, the performance of AmBC depends on the RF sources and strength of electromagnetic (EM) waves in the environment. As a potential way to change the EM properties, RIS can enhance AmBC from various aspects through adjusting propagation environments, such as performing interference management, expanding communication range, and providing additional high-quality links.

Although a number of surveys on this topic of IRC have recently appeared as \cite{3}\cite{4}, and references therein, it is still incompletely clear what roles can IRC technologies play in future IoE systems. Therefore, the main contribution of this article is to provide an in-depth understanding and discuss potential applications of IRC in future IoE systems as Fig. \ref{fig1}.

The rest of this paper is organized as follows. Section \uppercase\expandafter{\romannumeral2} explores two types of IRC-empowered communication scenarios in IoE. Then, Section \uppercase\expandafter{\romannumeral3} provides the mutual benefits of IRC-empowered IoE communication and sensing. Section \uppercase\expandafter{\romannumeral4} exposes the impacts of IRC on IoE security. Furthermore, Section \uppercase\expandafter{\romannumeral5} presents a summary of open issues and challenges in IRC-empowered IoE. Finally, Section \uppercase\expandafter{\romannumeral6} draws a conclusion.

\begin{figure*}[!t]
  \centering
  \includegraphics[width = 18cm,height= 7.5cm]{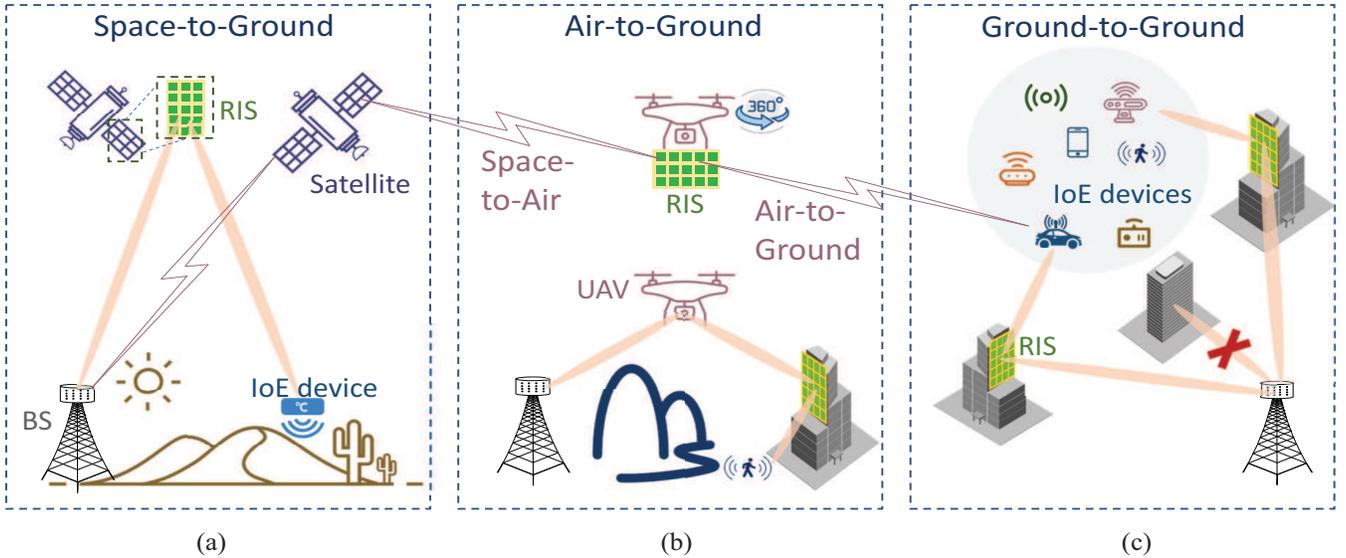}
  \caption{Architecture for RIS-empowered SAGIN in future IoE. (a) Deployment of RISs in satellite communications, (b) Deployment of RISs in aerial networks, and (c) Deployment of RISs on ground networks.}
  \label{fig2}
\end{figure*}

\section{Communication in IRC-Empowered IoE}
In this section, as shown in Fig. \ref{fig1}, we elaborate on the IRC-empowered IoE communication from two aspects, i.e., the space-air-ground integrated network (SAGIN) and green communications.
\vspace{-10.pt}
\subsection{Motivation of SAGIN and Green Communications}

The SAGIN is considered as one of the essential features for future wireless networks, which makes full use of low, medium and high frequency spectrum resources to achieve seamless global coverage \cite{5}. As an effective method to improve network coverage, the IRC technologies are naturally applied, together with satellite and UAV communications, to realize the vision of SAGIN. 

On the other hand, with widespread applications of IoE, a growing number of IoE devices are going to be connected. Hence, in practice, an important concern is to supply sufficient and sustainable power for all these devices. Although low-power wireless technologies enable battery life of current devices to last up one to three years, the labour cost of replacing batteries for an enormous number of devices is still unacceptable, especially for devices located in wilderness, culverts, tunnels, underwater, indoors, etc. Therefore, developing wireless technologies for green IoE communication becomes important, where IRC is envisioned to be a potential solution owing to its low power consumption.

\vspace{-10.pt}
\subsection{IRC-Empowered SAGIN in Future IoE}

\emph{1) Deployment of RISs in Satellite Networks:} For uninhabited areas where IoE devices distributed sparsely, e.g., deserts, oceans, and forests, low earth orbit (LEO) satellite communication has been an effective way of achieving seamless coverage, which is not susceptible to geographical factors. However, the LEO satellite brings the requirements of steerable antennas for tracking due to its motion with a relatively high velocity. Obviously, these requirements are unlikely to be met for IoE devices with low-complexity hardware. 

Specifically, the application of RIS in satellite networks is expected to deliver the benefits from two aspects. On one hand, RISs can reduce the hardware complexity of IoE devices by allowing over-the-air signal processing rather than requiring computations on the IoE devices. On the other hand, RISs are able to realize efficient beam tracking and beam directing by configuring the phase shift of each reflecting element, thus the coverage area is adaptively adjusted with respect to the stochastic geometry of IoE devices. A direct application of this case is to deploy RISs on the solar panel of LEO satellites for broadcasting and beamforming~\cite{6}, see Fig. 2(a), where the RISs' phase shifts are adjusted according to the IoE devices, as well as their propagation channels, to enhance the quality of service (QoS). Moreover, it is exemplified that an RIS-assisted satellite can provide up to an order-of-magnitude improvement in terms of the achievable rates \cite{6}.

Furthermore, high velocity of LEO satellites also brings challenges for localization especially when multiple interconnected LEO satellites cooperate to communicate with the IoE devices. To solve this, a state-of-the-art RIS based on the holographic principle becomes an attractive way of achieving high spatial resolutions in satellite networks. Specifically for the holographic RIS, a large number of tiny and inexpensive reconfigurable elements are integrated into a compact space in order to achieve holographic beamforming with a spatially continuous aperture \cite{7}. Compared to a conventional RIS, the holographic RIS generates desired beams through holographic recording and reconstruction so that higher spatial resolution is achieved. Meanwhile, it is easy to mount a holographic RIS on mobile terminals due to its highly integrated and ultra-thin structure. We envision that holographic RIS applied in LEO satellites can promote global connectivity and high-rate communications for IoE systems.

\emph{2) Deployment of RISs in Aerial Networks:} For complex terrains lacking infrastructure, e.g., mountain villages, temporary hotspots, and disaster areas, where the air-to-ground channels easily suffer from blockage, the combinations of unmanned aerial vehicles (UAVs) and RISs provide more flexibility and improve the signal quality of IoE devices in the absence of line-of-sight (LoS) connectivity \cite{8}. 

Fig. 2(b) depicts two typical cases of combining UAVs and RISs in aerial networks. In the first case, the RIS is installed on the facade of a building to reflect the signals from IoE devices to the UAV, which are further relayed to a nearby base station (BS). Virtual LoS links can be established between the UAV and IoE devices via RISs when direct LoS links are blocked by obstacles. Thus, the QoS of IoE devices can be stably met, which is beneficial to coverage extension. However, this deployment poses fundamental performance limitations. It can be difficult to find a suitable place for RIS installation and only the users at the same side of the RIS are served.

In a second case, RIS is mounted on an UAV (named as a flying RIS) to address the above issue \cite{8}. For example, an RIS is installed as a separate horizontal surface at the bottom of the UAV, which can be realized through lumped elements embedded in the reflector units. Compared to terrestrial RIS, the placement of the aerial RIS can be more flexibly optimized because of the mobility of the UAV and the ability of $360^{\circ}$ panoramic full-range coverage to serve a larger number of terminals. In addition, owing to the high altitude of the UAV, the flying RIS has a higher probability to establish strong LoS links with the IoE devices. 

\emph{3) Deployment of RISs on Ground Networks:} For urban areas where the IoE devices are distributed densely, the communication is vulnerable because the devices are frequently placed at cell edges or blocked by obstacles, where the transmitted signals suffer from deep fading. One solution to this problem is deploying RISs on outdoor buildings for providing supplementary links \cite{3}, as shown in Fig. 2(c). The RIS is composed of passive scattering elements with a designed physical structure. Each scattering element can be controlled in a software-defined manner to change the EM properties of the incident signals. By a joint control of all these artificial scattering elements, the reflecting phase shifts and angles of the incident signals can be properly tuned to create a desirable multi-path effect. Thus, the reflected signals are added constructively to improve the received signal power of the IoE devices. In addition, RIS can also be mounted in indoor ceilings to help reflect signals for extending the coverage when indoor users are located in dead zones.

Most current researches on terrestrial RISs focus on two-dimensional (2D) planar array. However, due to the limited coverage of terrestrial RIS compared to aerial RIS, terrestrial RIS can be extended to a 3D spherical surface \cite{9}. The spherical RIS exhibits additional advantages of higher received signal strength and wider coverage compared to flat RIS. It can thus incorporate with future networks comprising of devices from the ground vertically up to the space and bridge the terrestrial cellular networks, UAV networks, and satellite networks. Meanwhile, its influence on performance needs to be further investigated with beamforming optimization.

\begin{figure}[!t]
  \centering
  \includegraphics[width = 9cm,height= 7.5cm]{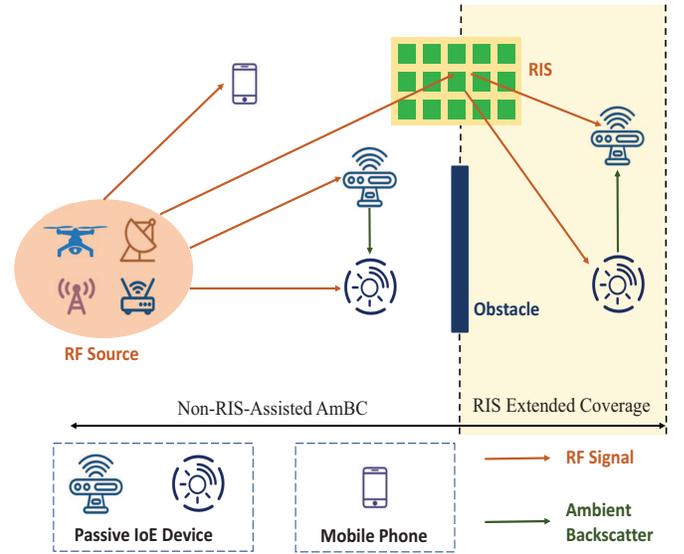}
  \caption{RIS-assisted AmBC communication systems.}
  \label{fig3}
\end{figure}

\renewcommand\arraystretch{2}
\begin{table*}[t]
  \caption{The potential communication scenarios for IRC-empowered IoE.}
  \begin{center}
    \begin{tabular}{>{\columncolor{pink!25}}l|lll}
      \hline
      \rowcolor{blue!15}\multicolumn{2}{l}{IRC-empowered IoE Schemes}&\multicolumn{1}{l}{Features}&\multicolumn{1}{l}{Benefits}\\
      \hline
      &\cellcolor{gray!20} &\cellcolor{gray!20}&$\bigcdot$ \cellcolor{gray!20}Enhance the QoS level\\
      &\cellcolor{gray!20}\multirow{-2}{*}{RIS-aided satellite networks}&\cellcolor{gray!20}\multirow{-2}{*}{$\bigcdot$ Mounted on the solar panel}&\cellcolor{gray!20}$\bigcdot$ Reduce hardware complexity of IoE devices\\
      &\multirow{2}{*}{RIS-aided UAV networks}&$\bigcdot$ Mounted on a fixed building&$\bigcdot$ Establish strong LoS links\\
      &&$\bigcdot$ Mounted on a mobile UAV&$\bigcdot$ Achieve $360^{\circ}$ panoramic full-range coverage\\
      \cline{2-4}
      &\cellcolor{gray!20}&\cellcolor{gray!20}$\bigcdot$ Mounted on outdoor buildings&\cellcolor{gray!20}$\bigcdot$ Create a desirable multi-path effect\\
      &\cellcolor{gray!20}\multirow{-2}{*}{RIS-aided ground networks}&\cellcolor{gray!20}$\bigcdot$ Mounted in indoor ceilings&\cellcolor{gray!20}$\bigcdot$ Extend the coverage of users in the dead zones\\
      \multirow{-7}{*}{Global Coverage in IoE}&AmBC&$\bigcdot$ Near various types of RF sources&$\bigcdot$ Energize passive IoE devices in SAGIN\\
      \hline
      &\cellcolor{gray!20}&\cellcolor{gray!20}$\bigcdot$ A combination of WIT and WPT&\cellcolor{gray!20}\\
      &\cellcolor{gray!20}\multirow{-2}{*}{RIS-aided SWIPT}&\cellcolor{gray!20}$\bigcdot$ Suitable for the downlink transmission&\cellcolor{gray!20}\multirow{-2}{*}{$\bigcdot$ Enhance energy harvesting performance}\\
      &RIS-aided AirComp&$\bigcdot$ Suitable for the uplink transmission&$\bigcdot$ Enhance wireless-powered data aggregation\\
      \cline{2-4}
      \multirow{-4}{*}{Energy efficiency in IoE}&\cellcolor{gray!20}AmBC&$\cellcolor{gray!20}\bigcdot$ Modulate information on reflected signals&\cellcolor{gray!20}$\bigcdot$ Enable passive devices for signal transmission\\
      \hline
    \end{tabular}
  \end{center}
  \label{tabel1}
\end{table*}

\emph{4) AmBC for SAGIN:} Different from RIS, AmBC allows passive IoE devices to harvest energy and to simultaneously transmit their information using the reflection of incident RF signals \cite{4}. As shown in Fig. \ref{fig3}, various types of signals in SAGIN can be utilized by AmBC, such as WiFi, cellular BSs, UAVs, satellites, etc. The WiFi signals are suitable for IoE deployment in indoor applications. The cellular BSs can be a preferred energy source in densely populated urban areas. Whereas the UAVs can be a supplement in sub-urban areas. Since the wide coverage characteristic, satellite signals can be a preferred solution for AmBC in uninhabited areas. In general, AmBC is an indeed promising communication technology to energize passive IoE devices in SAGIN.
\vspace{-10.pt}
\subsection{IRC-Empowered Green Communications in Future IoE}

\emph{1) RIS-Assisted Simultaneous Wireless Information and Power Transfer (SWIPT):} As a combination of wireless information transfer (WIT) and wireless power transfer (WPT), downlink SWIPT has been a viable technology to provide green communications with energy harvesting for ultra-massive low-power devices in IoE systems. However, energy receiver (ER) usually requires much higher power than an information decoder (ID). Therefore, how to improve the efficiency of ER is a critical issue for practical SWIPT systems.

As a cost-effective technology, RIS has the capability of achieving higher spectral and energy efficiencies. Motivated by the success of RIS in wireless communications, it is expected that RIS can play a promising role in SWIPT. In particular, by adjusting the phase shifts of each RIS element, the received signal at the ER can be strengthened, thereby improving the performance of SWIPT and empowering the envisioned battery-free IoE system. In addition, RIS-assisted mobile edge computing (MEC) is also a promising technique to assist SWIPT for IoE devices, by offloading computational tasks to the edge servers with reduced energy consumption \cite{10}.

\emph{2) RIS-Assisted Over-the-Air Computation (AirComp):} Different from downlink SWIPT, uplink AirComp has been developed to enable an efficient data-fusion of sensing data from many concurrent sensor transmissions by utilizing the inherent broadcast nature of wireless communications. AirComp achieves fast wireless data aggregation by reducing both computation and resource consumption, and efficiently power the IoE devices which are densely distributed.
However, in some harsh channel propagation conditions, the harvested energy of IoE devices can be too weak to support reliable uplink transmission, leading to high signal distortion in the AirComp. As a result, it is critical to mitigate these detrimental effects of channel fading. RIS is a promising solution to reconstruct the channel environment, which motivates researchers to design wireless-powered AirComp in RIS-aided IoE systems. Specifically, in \cite{11}, it reveals that deploying an RIS in AirComp can significantly reduce the mean-squared error, which validates the effectiveness of an RIS-aided AirComp.

\emph{3) AmBC for Green IoE:} AmBC technology plays an important role in realizing a green IoE communication system. Traditionally, IoE transceivers contain active RF components to generate and convert RF signals, which makes tiny IoE devices power hungry. Considering AmBC, it has the potential to supply sufficient energy to power tiny IoE devices, e.g., LEDs, pressure sensors, and accelerometers. By exploiting AmBC, the IoE devices do not need to generate their own RF signals, which avoids the requirement for power-consuming active components. It has been reported in \cite{12} that this helps reduce their energy consumption from milli-watts to micro-watts. Therefore, there is no need for dedicating power sources to these tiny IoE devices, since the micro-watts power requirement can be satisfactorily fulfilled by ambient RF signals.

Moreover, we stress that RIS and AmBC can be exploited cooperatively to enhance the communication in a future IoE. Fig.~\ref{fig3} depicts a novel system setup where AmBC devices operate in the presence of an RIS to minimize the consumption of transmit power. This combination further increases the coverage of IoE devices compared with non-RIS-assisted ones.

As a summary of this section, we list potential communication scenarios in IRC-empowered IoE in Table \ref{tabel1}, and illustrate the interplay of IRC and other technologies in Fig. \ref{fig1}.
\vspace{-10.pt}
\section{Sensing in IRC-Empowered IoE}
The use of high frequency bands from millimeter-wave (mmWave) to terahertz (THz), wide bandwidth, and massive antenna arrays in the future enables the integration of sensing and communication. The concept of ISAC is believed to be a key feature of future networks \cite{13}. In this section, as shown in Fig. \ref{fig1}, we analyze the IRC-assisted sensing and sensing-assisted IRC in future IoE systems, respectively, based on the architecture of ISAC.
\vspace{-10.pt}
\subsection{Enhancing Sensing by IRC for Future IoE}
Sensing enhancement is not easy under wireless connections, especially in future IoE systems. Specifically, it is impossible to consume extra resources to improve the quality of sensing since most of IoE devices are constrained nodes with limited power supply and computational capability. Hence, it is expected to use extra degrees of freedom to improve the performance of sensing and IRC is a promising candidate.
\begin{figure*}[!t]
  \centering
  \includegraphics[width = 18cm,height= 7cm]{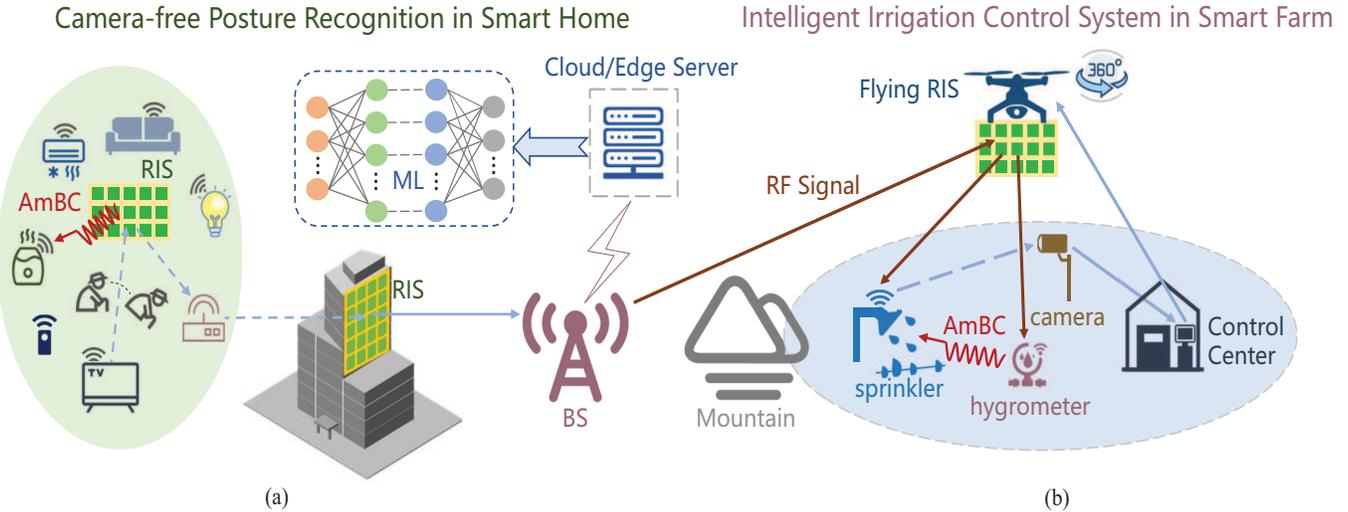}
  \caption{Potential scenarios with the integration of IRC and sensing for future IoE systems. (a) IRC enhances sensing in a smart home, and (b) Sensing enhances IRC in a smart farm.}
  \label{fig4}
\end{figure*}

Through the development of AmBC technology, smart sensors could be built, placed permanently inside a structure and set to communicate with each other to notify about this structure’s condition. The AmBC collects enough electricity to power sensors like LEDs, pressure sensors, accelerometers and so on. On the other hand, the communication links generated by RIS can be regarded as a large sensor \cite{13}. It explores the radio wave transmissions, reflections, and scatterings to obtain a better understanding of physical world. For instance, all IoE appliances such as televisions, doorbells, lights, and air conditioners are connected to the gateway in a smart home. However, indoor signals are easily blocked at high frequency bands, while an outdoor RIS installed on the buildings can help improve the communication links. In addition, indoor RISs can be installed on walls or ceilings, to increase the probability that the wireless channels are influenced by the sensing objectives. For example, Fig. 4(a) shows that posture recognition based on IRC-assisted sensing and machine learning (ML) in home IoE scenarios promotes camera-free supervision of, e.g., patients or elderly people, thus emergencies are identified immediately while individual privacy is protected. Even when the users are located in the dead zones where the direct receiving links are blocked, indoor RISs mounted on the ceilings helps reflect the signals for extending the range of sensing. Moreover, indoor WiFi signals can be further designed to power the passive IoE devices such as temperature/humidity sensors by AmBC.

\vspace{-8.pt}
\subsection{Enhancing IRC by Sensing for Future IoE} 
For IRC-assisted future IoE systems, most applications mentioned in articles are based on the assumption of perfect channel state information (CSI). However, it is always a challenge to obtain the CSI accurately in practice, especially when the channel dynamics are reconfigured by a passive device as RIS. Fortunately, sensing is able to help the IRC better understand the surrounding environment, thus facilitating efficient channel estimation and beam alignment. For example, in a system with obstacles between the BS and the IoE devices, the blockages in the direct LoS links can be sensed. After a success sensing, the beam is directly pointed to the RISs and then reflected to the IoE devices. It reduces unnecessary overhead and latency for beam recovery. Meanwhile, when the BS senses IoE devices at cell edges, it enhances the signal quality through transmitting sensing-assisted beams to nearby RISs.

Specifically, as shown in Fig. 4(b), sensing is also used to detect humidity levels for agricultural IoE applications, namely a smart farm. The hygrometer communicates with the sprinkler by using RF signals reflected by a flying RIS from the BS, where the AmBC is utilized to energize the passive IoE devices, i.e., a hygrometer and sprinkler. When the hygrometer detects that the humidity of the farmland is lower than a certain threshold, the sprinkler automatically discharges water. This linkage process relies on the quality of energy harvested from ambient wireless signals. A surveillance camera can upload information to the control center according to the perceived water discharge state of the sprinkler, and enhance the reflection of the ambient wireless signals by adjusting both trajectory and orientation of the flying RIS in real time. In summary, sensing can help provide a reliable and controllable communication link for IoE systems.

\vspace{-5.pt}
\section{Security in IRC-Empowered IoE}
In addition to the communication and sensing, security is of paramount importance, especially in the developing 5G networks as many IoE applications carry private, sensitive or confidential data. In recent years, incidents of IoE security occurred frequently, and important infrastructure, such as smart homes, traffic cameras, and even the national power grid, were attacked \cite{14}. Traditionally, security issues are addressed by complicated cryptographic solutions of high complexity, which are difficult to be applied in resource-constrained sensor devices in IoE. In recent years, physical layer security (PLS), by exploiting intrinsic characteristics of the communication medium as a complement to cryptographic methods, attracts increasing attentions \cite{15}. 

Due to the function of intelligent reflection, adoption of IRC can further make the PLS more efficient by involving dynamic channel environments into the design loop. In this section, we illustrate some typical potential applications of the IRC-enhanced PLS in IoE systems.

\vspace{-10.pt}
\subsection{Enhancing PLS in Unfavorable Secrecy Environments} 
In general, RISs can induce the required phase shifts on the reflected signals to maximize the signal-to-noise ratio (SNR) at the legitimate user, thus enhancing the performance of security, especially in the scenarios that are favorable for eavesdroppers.

\begin{figure}[!t]
  \centering
  \includegraphics[width = 9cm,height= 7cm]{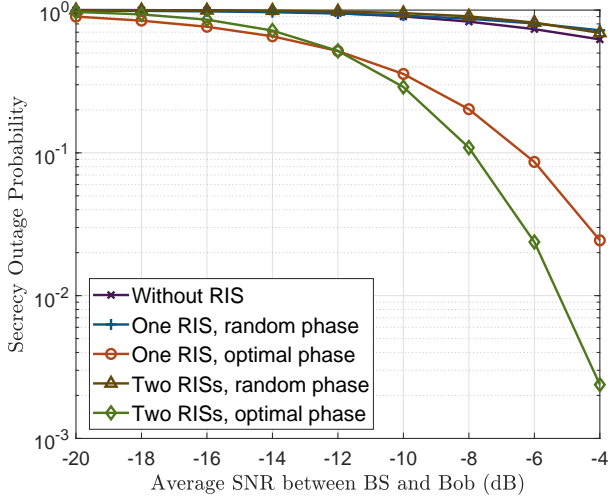}
  \caption{RIS-assisted IoE communication in unfavorable secrecy environments.}
  \label{fig5}
\end{figure}

As a case study, an RIS-assisted secure IoE communication system is considered, which is equipped with an RIS with sixteen elements. In this case, the link distance from the BS to the eavesdropper (Eve) is smaller than that from the BS to the legitimate IoE device (Bob). Specifically, the average SNRs of the BS-Eve and BS-RIS-Eve links are set to $-3~{\rm dB}$ and $0~{\rm dB}$, respectively. The average SNR of the BS-Bob link is equal to that of the BS-RISs-Bob link, and the target secrecy rate is $0.05$. To compare the secrecy performance of this efficient RIS-assisted IoE scheme, two baselines are additionally tested: the communication scheme without RIS and the communication scheme of RISs with random phase. From Fig. \ref{fig5}, the results show that the RISs with the optimal phase outperform the baseline schemes in terms of secrecy outage. The reason is that the RISs with the optimal phase realize accurate beamforming and maximally improve the received SNR at Bob. Moreover, Fig. \ref{fig5} also shows that the secrecy performance of the scheme with two RISs outperforms the scheme with a single RIS. This is because in the scenario of multiple RISs, more degrees of freedom are provided through the choices of the suitable RIS to maximize the received signal at the destination. In this way, the secrecy performance of Bob is further improved from the perspective of PLS.

\vspace{-10.pt}
\subsection{Enhancing PLS with Flying RISs} 
Flying RIS-assisted IoE communication has received much attention in recent years. Compared to a terrestrial RIS with a fixed location, the position of a flying RIS can be adjusted more flexibly in the 3D space with more refined beam resolutions for PLS enhancement. This feature improves the overall security by exploiting the high mobility of the flying RIS. 

Besides, it is envisioned that the flying RIS can serve as a mobile cooperative jammer jointly with the ground BS to improve the security. Through transmitting the artificial noise (AN) signals to the malicious nodes, the drawbacks of low hardware complexity for IoE devices to secure communication can be overcame by the flying RIS.

Furthermore, in a scenario where multiple Bobs and Eves exist, a single flying RIS has to fly over some Eves due to its mission requirement, where the possibility of information leakage bursts. This motivates the deployment of multiple collaborative flying RISs to achieve stronger secure communications. For example, based on the locations of Eves, the ground Bobs can be grouped into different clusters with each cluster served by a single flying RIS. In specific, a part of the flying RISs can act as aerial jammers above the Eves, while the others are deployed near Bobs. As such, it may not be necessary for the flying RIS to fly over the Eves, which thus helps reduce the information leakage.

\vspace{-1.pt}
\section{Open Issues and Challenges}
The IRC-empowered future IoE is a growing research area. Addressing the associated design issues is the key to unlock its potential. Despite fruitful research in this area, there are a variety of challenges waiting to be tackled. In this section, we highlight open issues and challenges for future research.
\vspace{-10.pt}
\subsection{Joint Design of Flying RIS for SAGIN}
The flying RIS faces multiple technical challenges for the joint design. The setups of the UAV and RIS should be taken into account together with the transmission optimization. First for UAV, higher requirements of endurance and controllability should be satisfied when carrying RIS. Besides, its trajectory needs to be jointly optimized with the transmit beamforming at the BS and the passive beamforming at RIS, to maximize the performance metrics (e.g., achievable rate, spectral efficiency, ergodic capacity, etc.) of terrestrial nodes. Secondly for RIS, we need to consider multiple factors such as size, orientation and placement when deploying it on the UAV. It is noteworthy that the orientation of RIS mounted on the UAV can significantly affect the air-ground channels.
\vspace{-10.pt}
\subsection{Dynamic CSI Acquisition} 
It is observed that a majority of IRC applications are based on the assumption of perfect CSI at the transmitter or the IRC devices. In practice, both RISs and AmBC are passive, and the channel estimation by low-power IoE devices is a challenging task. Meanwhile, the computational complexity of channel estimation grows rapidly as the number of RIS reflection elements increases. In addition, Eves may attack the channel estimation by sending the same pilot sequence, which further deteriorates the channel estimation.
\vspace{-10.pt}
\subsection{Deployment of RISs and AmBC}
Generally speaking, the deployment of RISs at different locations is an intractable problem compared to the deployment of BSs or relays because of the passive nature of RISs. It is still an open challenge to adjust the physical design, deployment, collaboration, and association to enhance RISs-assisted IoE. ML and stochastic geometry-based solutions may be good options for efficient deployment of RISs.

AmBC typically covers an area ranging from several to tens of meters with low data rates. There is an immense need for improvement in range extension, to enhance the coverage of AmBC-assisted IoE. In addition, simultaneous transmission of many passive devices can create uncontrollable interference. Interference management by using AmBC among ultra-massive IoE devices becomes of increasing interest.

\vspace{-10.pt}
\section{Conclusion}
This article provides a comprehensive survey on the applications of IRC in future IoE systems. Specifically, RIS and AmBC are two different forms of IRC. First, the potential benefits in the context of SAGIN and green communications are revealed. RISs and AmBC can be exploited cooperatively to enhance the communication performance of future IoE in these two scenarios. Then, we also analyze the integration of RF sensing and IRC technologies in IoE. IRC technologies can improve sensing performance, and accurate sensing information can provide more reliable communication links. Moreover, we propose to apply IRC to enhance PLS in unfavorable secrecy environments for IoE. Research indicates that flying RISs have great potential in PLS enhancement. Open issues and challenges for realizing IRC-assisted future IoE systems are identified in terms of the joint design of flying RIS for SAGIN, dynamic CSI acquisition, and the deployment of IRC devices.

\vspace{-15.pt}
\section*{Acknowledgment}
This work was supported in part by the National Key Research and Development Program of China under Grant 2020YFB1806600; the National Natural Science Foundation of China (NSFC) under Grants 62022026 and 6211101429; and the Fundamental Research Funds for the Central Universities under Grant 2242022k30002.

\vspace{-10.pt}
\section*{Biographies}

\textsc{Wei Shi} (wshi@seu.edu.cn) received the B.S. degree in communication engineering from Nanjing University, Nanjing, China, in 2017. In 2019, he received his M.S. degree in information and communication engineering from Southeast University, Nanjing, China. He is currently working towards the Ph.D. degree in information and communication engineering at the National Mobile Communications Research Laboratory, Southeast University, Nanjing, China. His current research includes intelligent reflection surface and physical layer security.

\textsc{Wei Xu} (corresponding author, wxu@seu.edu.cn) received the Ph.D. degree from Southeast University, Nanjing, China in 2009. From 2009 to 2010, he was a Post-Doctoral Research Fellow at the Department of Electrical and Computer Engineering, University of Victoria, Canada. He is currently a Professor at the National Mobile Communications Research Laboratory, Southeast University. He is currently an Editor of IEEE Transactions on Communications, and also a Senior Editor of IEEE Communications Letters.

\textsc{Xiaohu You} (xhyou@seu.edu.cn) received his Ph.D. degree from Nanjing Institute of Technology in 1989. He is with Southeast University, Nanjing, China, where he has been a professor since 1990. His research interests include mobile communications, adaptive signal processing, and artificial neural networks. He was the recipient of the National First Class Invention Prize in 2011. He was selected as an IEEE Fellow in 2012 for his contributions to the development of mobile communications in China.

\textsc{Chunming Zhao} (cmzhao@seu.edu.cn) received the Ph.D. degree from the Department of Electrical and Electronic University of Kaiserslautern, Germany, in 1993. He has been a Professor and the Vice Director of the National Mobile Communications Research Laboratory, Southeast University. He has managed several key projects of the Chinese Communications High Technology Program. His research interests include communication theory, coding/decoding, and VLSI design. He won the First Prize of National Technique Invention of China in 2011.

\textsc{Kejun Wei} (weikejun@caict.ac.cn) received the Ph.D. degree from the PLA University of Science and Technology, Nanjing, China, in 2004. From 2005 to 2009, he was a post-Doctoral Researcher with the Institute of Acoustics, Chinese Academy of Sciences. He is currently a Senior Engineer with the China Academy of Information and Communications Technology. He deeply participates in the research activities of the IMT-2020 (5G) Promotion Group and IMT-2030 (6G) Promotion Group.

\end{document}